\documentclass[traditabstract]{aa}

\usepackage{txfonts}
\usepackage{epsf}
\usepackage{graphicx}
\usepackage{color}
\usepackage{natbib}

\newcommand{\ten}[2]{#1\times 10^{#2}}
\newcommand{\atoms}{H-atoms\,cm$^{-2}$}
\newcommand{\gs}{g\,s$^{-1}$}
\newcommand{\erg}{erg\,s$^{-1}$}
\newcommand{\ergs}{erg\,cm$^{-2}$s$^{-1}$}
\newcommand{\ergsa}{erg\,cm$^{-2}$s$^{-1}$\AA$^{-1}$}

\newcommand{\cts}{$\rm {cts\,s}^{-1}$}
\newcommand{\kms}{$\rm {km\,s}^{-1}$}
\newcommand{\msun}{$M_\odot$}

\newcommand{\RX}{RX~J1007.5--2017}
\newcommand{\rxj}{RXJ1007}

\newcommand{\ha}{H$\alpha$}
\newcommand{\hb}{H$\beta$}
\newcommand{\hg}{H$\gamma$}
\newcommand{\heps}{H$\epsilon$}
\newcommand{\oc}{$O\!-\!C$}

\begin{document}

\title {The high-field polar RX~J1007.5--2017}

\author
   {H.-C. Thomas\inst{1}\thanks{Deceased 18 January 2012}
      \and K. Beuermann\inst{2}
      \and K. Reinsch\inst{2}
      \and A.D. Schwope\inst{3}
      \and V. Burwitz\inst{4}
   }


\institute{
MPI f\"ur Astrophysik, Karl-Schwarzschild-Str. 1, D-85740 Garching, Germany
\and Institut f\"ur Astrophysik, Friedrich-Hund-Platz 1, 37077 G\"ottingen, Germany,
\and Leibniz-Institut f\"ur Astrophysik Potsdam (AIP), An der Sternwarte 16, 14482 Potsdam, Germany
\and MPI f\"ur extraterrestrische Physik, Giessenbachstr. 6, 85740 Garching, Germany
   }

\date{Received 25 June 2012 / Accepted 3 September 2012}

\titlerunning{The high-field polar RX~J1007.6--2017}
\authorrunning{H.-C. Thomas et al.}

\abstract 
{We report optical and X-ray observations of the high-field polar
  \RX{} performed between 1990 and 2012. It has an orbital period of
  208.60~min determined from the ellipsoidal modulation of the
  secondary star in an extended low state. The spectral flux of the
  dM3- secondary star yields a distance of $790\pm 105$\,pc. At low
  accretion levels, \RX{} exhibits pronounced cyclotron emission
  lines. The second and third harmonic fall in the optical regime and
  yield a field strength in the accretion spot of 94\,MG.  The source
  is highly variable on a year-to-year basis and was encountered at
  visual magnitudes between $V\!\sim\!20$ and $V\!\sim\!16$. In the
  intermediate state of 1992 and 2000, the soft X-ray luminosity
  exceeds the sum of the luminosities of the cyclotron source, the hard
  X-ray source, and the accretion stream by an order of magnitude. An X-ray
  high state, corresponding to the brightest optical level, has
  apparently not been observed so far.}
{}{}{}{}

 \keywords {stars: cataclysmic variables -- stars: magnetic fields --
   stars: individual: RX~J1007.5--2016 -- binaries: close -- X-rays:
   stars -- accretion } 

\maketitle 

\section{Introduction} 

In polars, a subclass of cataclysmic variables (CVs), a magnetic white
dwarf accretes matter from its Roche-lobe filling low-mass
main-sequence companion. The strong magnetic field channels the flow
onto a small accretion spot on the surface of the white dwarf, where
the energy not expanded in the accretion stream is released in form of
hard X-rays, soft X-rays, and cyclotron radiation. About 100 of these
systems are known, most of them discovered as bright soft X-ray and
EUV sources. The spectral characteristics of the spot emission depend
sensitively on the magnetic field strength $B$ in the accretion spot
and on the density of the accreted matter (e.g.,
\citep[e.g.][]{kuijperspringle,fischerbeuermann}. The distribution of
observed field strengths in polars ranges from 10\,MG to about 200 MG,
with only three stars known to possess a field strength exceeding
100\,MG \citep{schwope09}.  For comparison, single magnetic white
dwarfs are known to possess field strengths up to 1000\,MG and display
a higher incidence of very high field strengths. It is not known why
such systems are lacking among the polars. Either post-common envelope
binaries contain fewer white dwarfs with very high field strengths, or
we fail to detect these systems once they reach the CV-stage. \RX\
(henceforth \rxj) is one of the few high-field polars. We present here
a comprehensive analysis of its properties.

\begin {table}[b]
\caption[]{Summary of observations: X = X-ray, S
  = spectrophotometry, C = circular spectropolarimetry, P =
  photometry, WL = white light.}
\label{tab:observations}
\begin {flushleft}
\begin {tabular}{r@{\hspace{1mm}}l@{\hspace{0mm}}l@{\hspace{3mm}}l@{\hspace{2mm}}l@{\hspace{2mm}}l@{\hspace{2.5mm}}l}
\hline
\noalign{\smallskip}
\multicolumn{3}{l}{\hspace{9.0mm}Date} & Telescope/Instrument & & Range & Resol.\\
\noalign{\smallskip}
\hline
\noalign{\smallskip}
24--26 & Nov & ~1990 & ROSAT\,/\,PSPC & X & 0.11--2.0\,keV & \\
 9--13 & Jan & ~1992 & ESO 2.2m\,/\,EFOSC2 & S & 3500--9000\AA  & 40\,\AA\\
       &     &       & & S & 3500--5400\AA  & 10\,\AA\\
17 & Nov & ~1992 & ROSAT\,/\,PSPC & X & 0.11--2.0\,keV & \\
15 & Nov & ~1995 & ESO 2.2m\,/\,EFOSC2 & S & 3800--9100\AA  & 30\,\AA\\
6 & Mar & ~1997 & ESO 2.2m\,/\,EFOSC2 & S & 3500--10200\AA  & 50\,\AA\\
14 & Dec &~1999 & 2MASS & P & J, H, K$_\mathrm{s}$  & \\
2--3 & Feb & ~2000 & ESO 3.6m\,/\,EFOSC2 & C & 3378--7516\AA  & 15\,\AA\\
7 & Dec & ~2001 & XMM\,/\,EPIC pn & X & 0.22--8.5\,keV & \\
19--20 & Jan & ~2010 & MONET/N          & P & I$_\mathrm{Bessell}$, Sloan g  & \\ 
Jan~-- & Mar & ~2011 & MONET/N          & P & I$_\mathrm{Bessell}$, WL & \\ 
2 & Jan & ~2012 & MONET/N          & P & I$_\mathrm{Bessell}$  & \\ 
\noalign{\smallskip}
\hline
\end {tabular}
\end {flushleft}
\end {table}

\rxj\ was discovered in 1990 as a bright, very soft X-ray source in
the ROSAT All Sky Survey and was subsequently identified with a
19-mag magnetic cataclysmic variable of the polar subclass
\citep{bt93,tb98,reinschetal99}.
In this paper, we summarize our observations of \rxj\ carried out over
more than two decades, from its discovery in 1990 to 2012. The
analysis of the early data suffered from the lack of a sufficiently
accurate ephemeris, preventing a proper phasing of the data taken over
the years in different wavelength regimes. This problem has now been
overcome and the entire set of so-far unpublished data is presented
here. Our analysis is based on optical observations, comprising
time-resolved spectrophotometry, spectropolarimetry, and
photometry. It includes an analysis of the ROSAT X-ray data and a
re-analysis of the XMM-Newton X-ray data previously discussed by
\citet{ramsaycropper03}.

\section{Observations and data analysis} 

Table~\ref{tab:observations} contains a log of our own and previously
published observations. \rxj\ displayed substantial long-term
variability both at X-ray and optical wavelengths. It was encountered
in low states of accretion in 1997 and 2011/2012, in intermediate
states in 1992, 2001, and 2010, and in a high state in 2000.  The
optical position of \rxj\ is $\alpha_{2000}=10^{\rm h} 07^{\rm m}
34\fs 6$, $\delta_{2000}=-20^\circ 17' 32\arcsec$.

\subsection{X-ray observations}

\rxj\ was observed from 24--26 November 1990 in the ROSAT
All-Sky-Survey (RASS) with the Position Sensitive Proportional Counter (PSPC)
as detector for a total of 510\,s (25 sightings). The highly variable
source 1RXS J100734.4-201731 had a mean count rate of
1.0\,\cts\ and a very soft spectrum.
It was subsequently observed with the ROSAT PSPC on 17 November 1992
with a total exposure time of 10\,ks and again found to be very soft
with a lower mean count rate of 0.37\,\cts. \citet{ramsaycropper03}
observed the source on 7 December 2001 with XMM-Newton with the
European Photon Imaging Camera (EPIC pn) for 5.5\,ks. This observation
showed the source in a brighter state again and confirmed the very
soft X-ray spectrum.

\begin{figure}[t]
  \includegraphics[height=8.9cm,bb=52 61 543 700,angle=270,clip=]{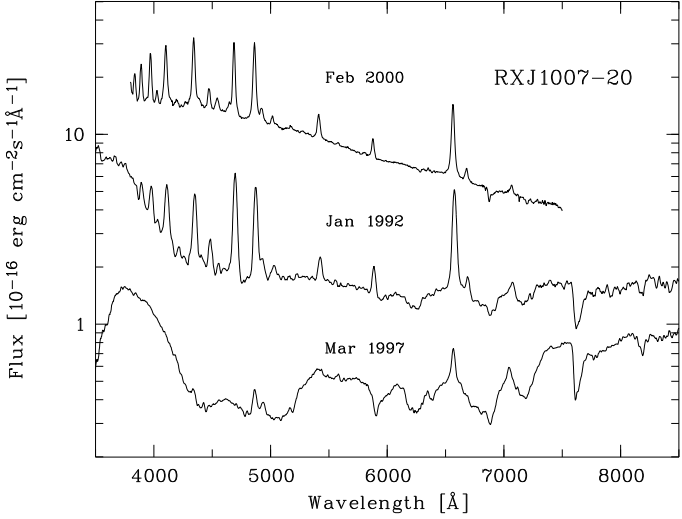}
  \caption{Mean spectra of \rxj\ in February 2000 (top), January 1992
    (center), and March 1997 (bottom). }
\label{meanspectra} 
\end{figure}

\begin {table}[b]
\caption[]{New photometric minima with \oc\ residuals from Eq.~1. }
\label{tab:monet}
\begin {flushleft}
\begin {tabular}{@{\hspace{3mm}}r@{\hspace{4.2mm}}c@{\hspace{4.2mm}}c@{\hspace{4.2mm}}r@{\hspace{4.2mm}}rl}
\hline
\noalign{\smallskip}
E\hspace{3.5mm} & BJD(TDB) & Error & O--C\hspace{1mm} & Orbital & Band \\
& 24000000+ & (day) & (day)\hspace{0.7mm} & Phase\hspace{0.6mm} & \\
\noalign{\smallskip}
\hline
\noalign{\smallskip}
 --45458.0 & 48630.7399 &  0.0020 &   0.0015 &    0.010 & \hspace{-1.0mm}7500\AA \\
 --32460.0 & 50513.6773 &  0.0020 &$-$0.0023 & $-$0.016 & \hspace{-1.0mm}7500\AA \\
      0.0 & 55215.9649 &   0.0020 &   0.0023 &    0.016 & $I_\mathrm{Bessell}$ \\
     --0.5 & 55215.8915 &  0.0020 &   0.0014 &    0.501 & $I_\mathrm{Bessell}$ \\
      6.0 & 55216.8337 &   0.0030 &   0.0020 &    0.014 & Sloan g \\
   2408.5 & 55564.8625 &   0.0030 &$-$0.0048 &    0.467 & WL \\
   2409.0 & 55564.9378 &   0.0030 &$-$0.0019 & $-$0.013 & WL \\
   2415.5 & 55565.8790 &   0.0020 &$-$0.0024 &    0.483 & WL \\
   2457.0 & 55571.8921 &   0.0020 &$-$0.0011 & $-$0.008 & WL \\
   2457.5 & 55571.9622 &   0.0020 &$-$0.0034 &    0.477 & WL \\
   2484.5 & 55575.8754 &   0.0020 &$-$0.0016 &    0.489 & $I_\mathrm{Bessell}$ \\
   2485.0 & 55575.9514 &   0.0020 &   0.0020 &    0.014 & $I_\mathrm{Bessell}$ \\
   2485.5 & 55576.0200 &   0.0030 &$-$0.0019 &    0.487 & $I_\mathrm{Bessell}$ \\
   2587.5 & 55590.7981 &   0.0020 &   0.0002 &    0.501 & $I_\mathrm{Bessell}$ \\
   2588.0 & 55590.8689 &   0.0040 &$-$0.0015 & $-$0.011 & $I_\mathrm{Bessell}$ \\
   2588.5 & 55590.9487 &   0.0060 &   0.0059 &    0.541 & $I_\mathrm{Bessell}$ \\
   2594.5 & 55591.8151 &   0.0020 &   0.0031 &    0.522 & WL \\
   2595.0 & 55591.8857 &   0.0020 &   0.0013 &    0.009 & WL \\
   2642.5 & 55598.7691 &   0.0030 &   0.0036 &    0.525 & $I_\mathrm{Bessell}$ \\
   2643.0 & 55598.8402 &   0.0020 &   0.0023 &    0.016 & $I_\mathrm{Bessell}$ \\
   2643.5 & 55598.9086 &   0.0030 &$-$0.0017 &    0.488 & $I_\mathrm{Bessell}$ \\
   2835.5 & 55626.7192 &   0.0050 &$-$0.0050 &    0.466 & $I_\mathrm{Bessell}$ \\
   2836.0 & 55626.7961 &   0.0030 &$-$0.0006 & $-$0.004 & $I_\mathrm{Bessell}$ \\
   2843.0 & 55627.8101 &   0.0030 &$-$0.0006 & $-$0.004 & WL \\
   4921.5 & 55928.9102 &   0.0020 &$-$0.0001 &    0.499 & $I_\mathrm{Bessell}$ \\
\noalign{\smallskip}
\hline
\end {tabular}
\end {flushleft}
\end {table}

\subsection{Optical spectrophotometry and spectropolarimetry}

Phase-resolved spectrophotometry was performed in January 1992, using
the ESO/MPI 2.2-m telescope with EFOSC2 and either grism G1 (FWHM
resolution 50\,\AA) or grism G3 (FWHM resolution 10\,\AA). Further
low-resolution spectrophotometry was performed on 6~March 1997.  On
3~February 2000, when the source was at the brightest level
encountered so far, we performed circular spectropolarimetry, using
the ESO 3.6-m telescope with EFOSC2 and a customer-supplied
quarter-wave plate. All spectrophotometry was placed on an absolute
flux scale using standard stars. Fig.~\ref{meanspectra} shows the mean
spectra of \rxj\ in the 1992, 1997, and 2000 observations, which
demonstrate its variability. In the 1997 low state, the red part of
the spectrum is dominated by the TiO bands of the secondary star and
the visual magnitude varies over the orbit between
$V\!=\!19.4-20.0$. The weak Balmer and the intense cyclotron emission
lines indicate that low-level accretion is still taking place. The
1992 intermediate and 2000 high states are characterized by
significantly increased levels of Balmer line emission and the
associated continuum, but only by a moderate increase in cyclotron
emission, as discussed in Sect.~\ref{sec:cyc}. In the 2000 high state,
the source varied over the orbit between $V\!=\!16$ and 17.

The profiles of the Balmer and helium emission lines in the
medium-resolution spectra of 1992 and 2000 display phase-dependent
asymmetries, but the narrow and broad line components, typical of
polars, cannot reliably be separated. For simplicity, we measured
radial velocities by fitting single Gaussians to the strong Balmer and
HeII$\lambda$4686 emission lines. Given the lack of better-resolved
low-state spectra, we measured radial velocities and line fluxes also
from the weak H$\alpha$ emission lines as well as radial velocities
from the unresolved near-infrared NaI$\lambda 8190$ absorption-line
doublet in the 1997 spectra. In retrospect, the phasing of the
velocities suggests that the bulk of the intermediate and high-state
emission originates from the magnetically confined part of the
accretion stream and the low-state Balmer emission from the heated
face of the secondary star.

\begin{figure}[t]
  \includegraphics[width=86.5mm,viewport=54 112 482 760,clip=]{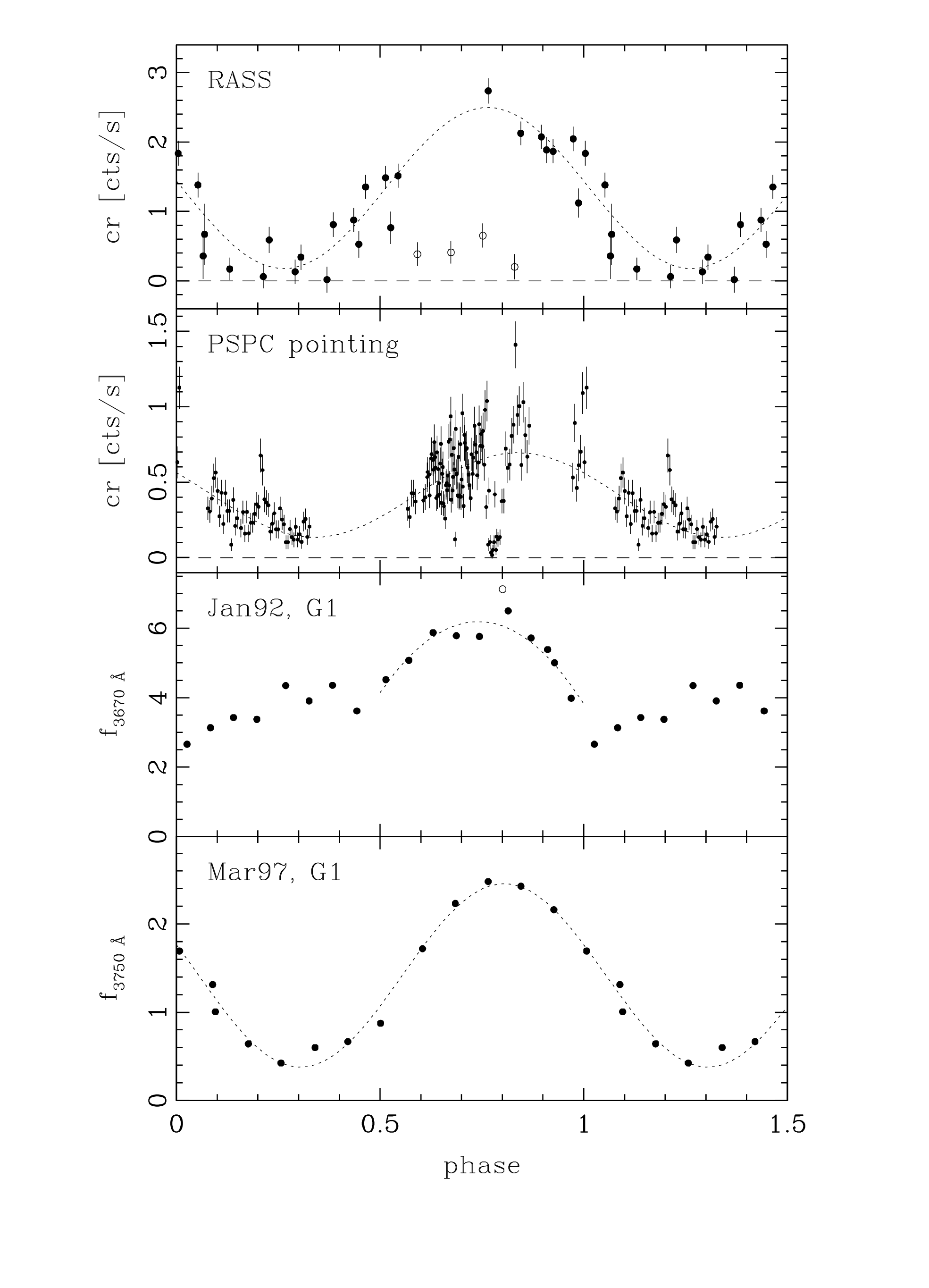}

\vspace*{-2.8mm}
  \includegraphics[height=88.7mm,bb= 76 65 558 714,angle=270,clip=]{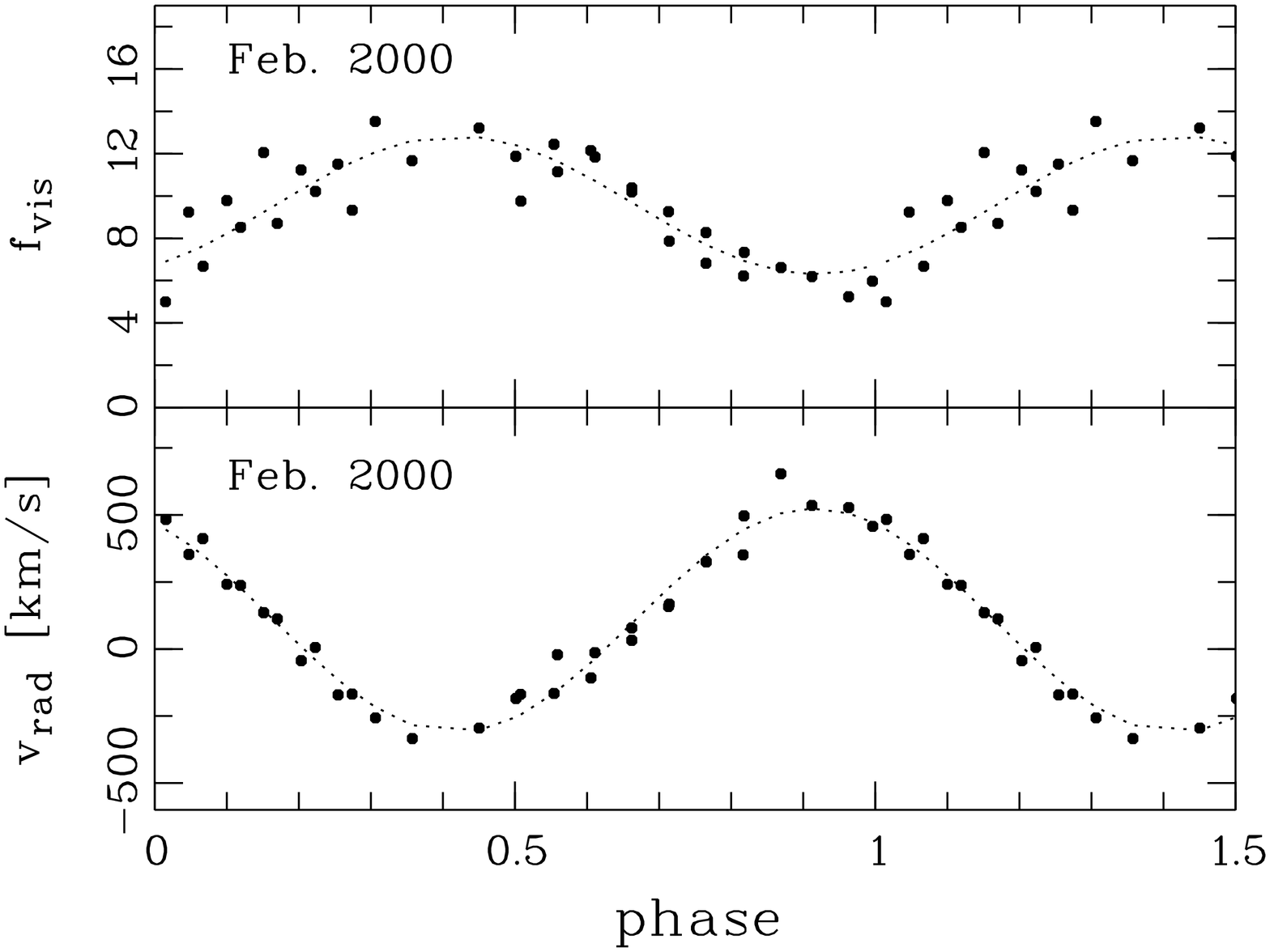}
  \caption{Orbital variations of \rxj\ from top: (1) 1990 X-ray light
    curve from the ROSAT All Sky Survey; (2) 1992 ROSAT pointed
    observation; (3) 1992 quasi $U$-band flux; (4) 1997 quasi $U$-band
    flux; (5) 2000 visual flux; (6) 2000 Balmer/HeII emission-line radial 
velocity. The optical fluxes are in units of $10^{-16}$\,\ergsa.}
\label{lc1} 
\end{figure}

\begin{figure}[t]
  \includegraphics[height=88.7mm,bb= 91 67 502 684,angle=270,clip=]{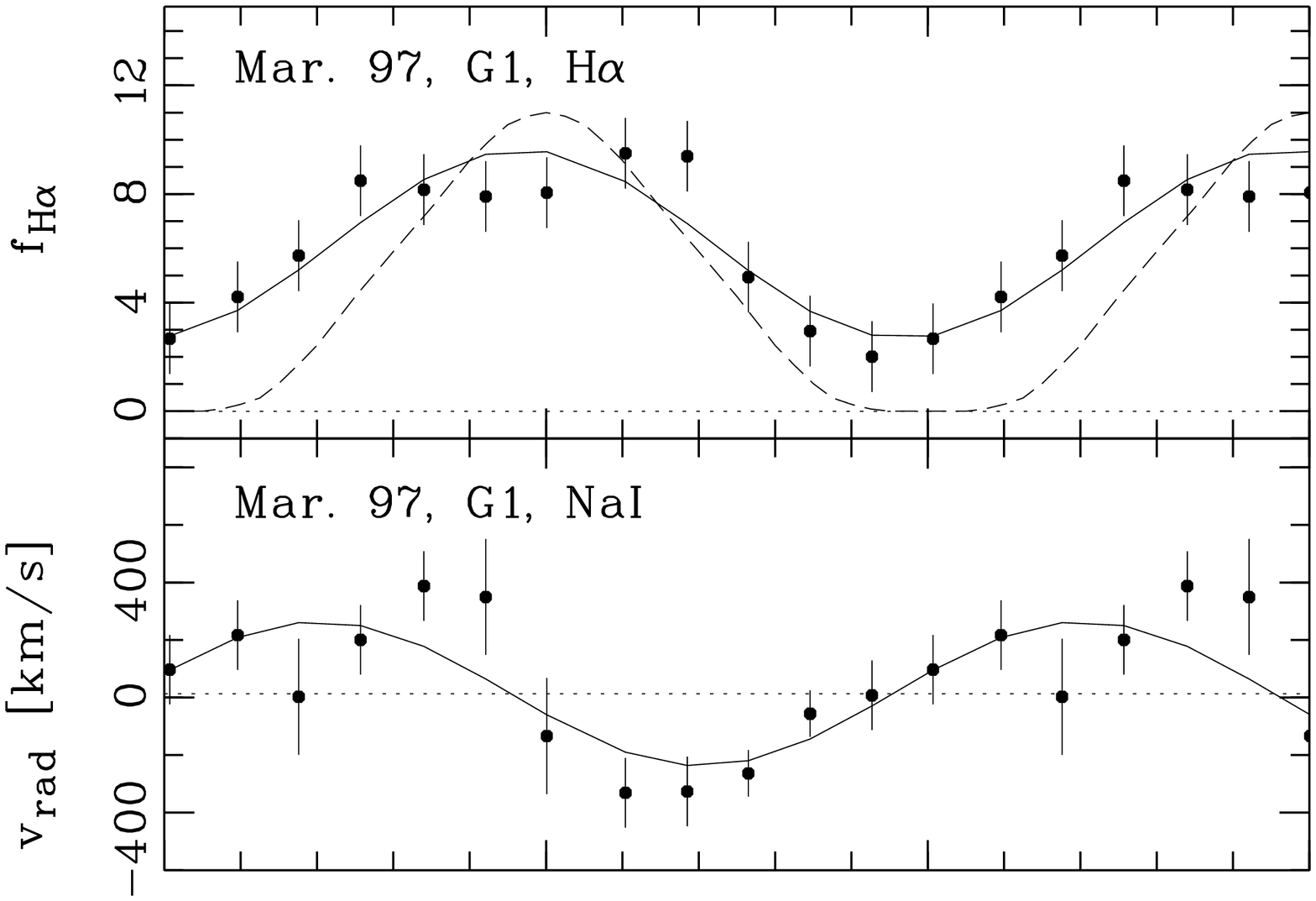}

\vspace*{-0.8mm}
  \includegraphics[width=89.5mm,viewport=51 60 493 746,clip=]{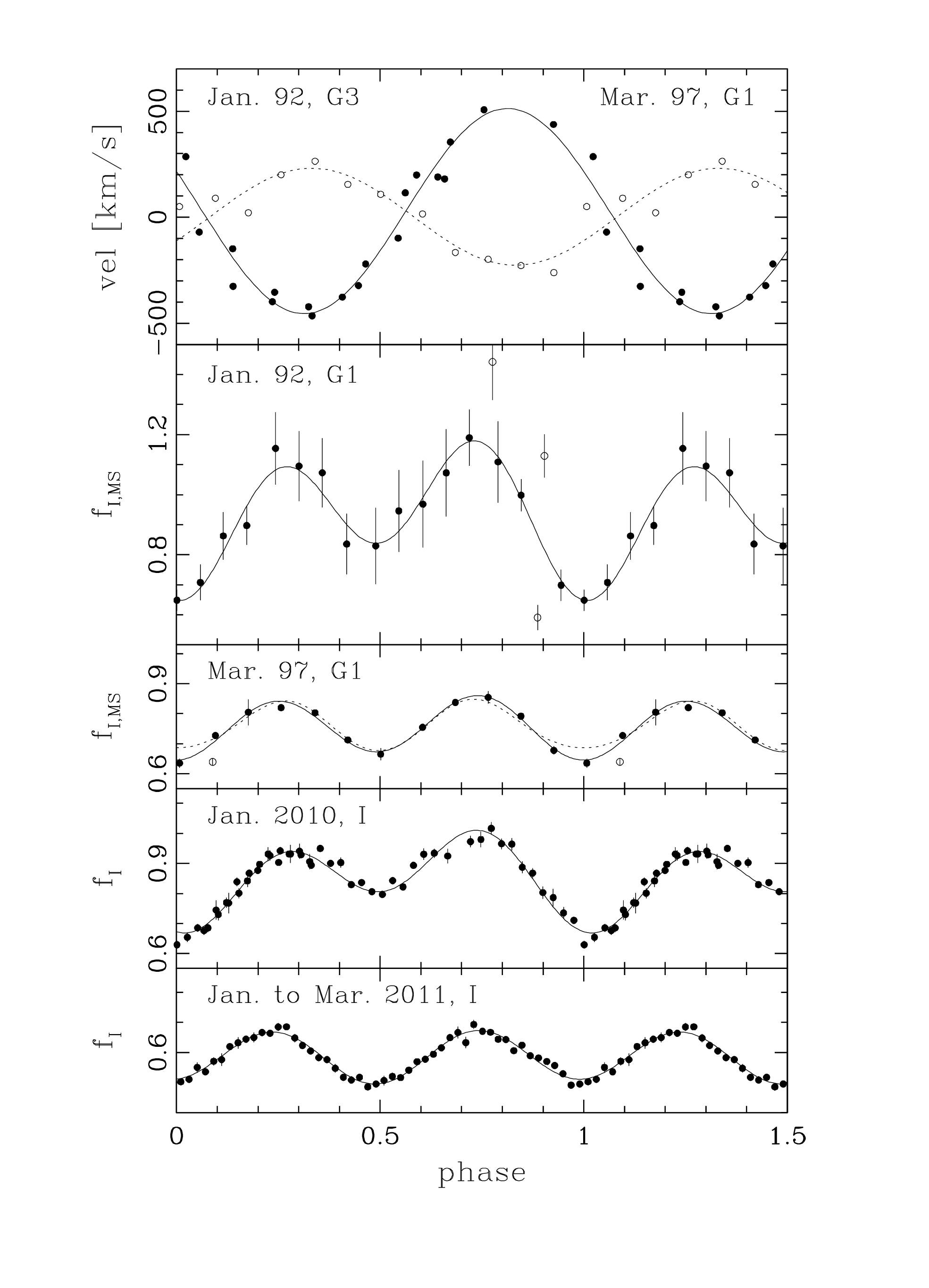}
  \caption{Orbital variations continued: (1) 1997 \ha\ emission-line
    flux; (2) 1997 NaI$\lambda8190$ absorption-line
    radial velocity ; (3) 1997 \ha\ emission-line radial velocity
    (open circles, dashed curve) and 1992 Balmer/HeII emission-line
    radial velocity (filled circles, solid curve); (4) 1992
    $f_\mathrm{7500\AA}$ flux; (5) 1997$f_\mathrm{7500\AA}$ flux; (6)
    2010 $I$-band flux; (7) 2011 $I$-band flux.}
\label{lc2} 
\end{figure}

\subsection{Optical and IR photometry}

We performed time-resolved relative optical photometry in 2010--2012,
using the MONET/North telescope at McDonald Observatory via the MONET
browser-based remote-observing interface. The data were taken with an
Apogee ALTA E47+ 1k$\times$1k CCD camera mostly in Bessel $I$ (central
wavelength 8000\,\AA) and white light (clear filter) with exposure
times of 60\,s. Photometry was performed relative to the comparison
star USNOA 0675\,10804546 (2000:
RA=$10^{\mathrm{h}}07^{\mathrm{m}}32.87^{\mathrm{s}}$,
DEC=$-20^\circ15`36.3"$), which has a DENIS $I$-band magnitude of
13.14$\pm$0.03 \citep{fouqueetal00}. Combined with standard-star
measurements in the 1992 and 1997 campaigns, we derived a common
$I$-band calibration for the spectrophotometry and photometry.
All our $I$-band data display a clear double-wave orbital modulation.
\rxj\ was found at $I\!=\!17.5\!-\!18.1$ in January 1992,
$I\!=\!17.75\!-\!18.08$ in March 1997, $I\!=\!17.28\!-\!17.74$ in
January 2010, and $I\!=\!17.71\!-\!18.08$ in January--March 2011 as
well as January 2012. 
The source has an entry in the 2-micron All Sky Survey (2MASS)
catalog, which gives $J\!=\!16.30\!\pm\!0.10, H\!=\!15.63\!\pm\!0.12$,
and $K_\mathrm{s}\!=\!15.22\!\pm\!0.13$.

\section{Results}

In this section, we describe first the effort to derive an accurate
orbital ephemeris, which allows us to correctly phase the
observations collected over more than 20 years, and then continue to
discuss  the physical properties of the system.

\subsection{Orbital period}
\label{period}

Early in the project, we obtained approximate values of the orbital
period $P$ of \rxj\ by folding the X-ray fluxes (Fig.~\ref{lc1},
panels 1 and 2 from top), the optical spectrophotometric fluxes, and
the Balmer and HeII emission-line radial velocities (Fig.~\ref{lc1},
panels 3--6, and Fig.~\ref{lc2} panel 3 from top, solid circles). The
individual observations yielded $P\!=\!208.1\!\pm\!0.7$\,min
(RASS), $P\!=\!205\!\pm\!3$\,{\bf min} (ROSAT pointed), and
$P\!=\!208.2\!\pm\!0.3$\,min (1992 radial velocities),
suggesting $P\!\simeq\!0.145$\,d. All attempts to derive a long-term
ephemeris from these data suffered from alias problems.

An alternative approach is suggested by the quasi-$I$ band fluxes in
Fig.~\ref{lc2} (panels 4 and 5 from top), which show what may be
the ellipsoidal modulation of the secondary star in the 1992
intermediate and the 1997 low-state data. The two light curves
refer to the 7450--7550\,\AA\ band, which measures the flux of the
secondary star with some contribution from the white dwarf and the
accretion stream. These early data were supplemented by the $I$-band
light curves taken in the 2010 intermediate and the 2011--2012
extended low states (Fig.~\ref{lc2}, bottom two panels). For the dM3--
secondary star in \rxj\ (Sect.~\ref{sec:sec}), the spectral flux at
7500\,\AA\ and the mean Bessell $I$-band flux agree within a few
percent (Beuermann 2006), allowing an easy comparison. The four light
curves shown have a common ordinate scale normalized to the peak flux
in 2010, which corresponds to $I\!=\!17.3$. Standard theory of the
ellipsoidal modulation for Roche-lobe filling stars predicts two
minima, of which the deeper one, according to von Zeipel's (1924) law,
occurs at superior conjunction of the secondary star. This order may
be reversed if the low-gravity hemisphere of the secondary is
radiatively heated by the accreting white dwarf. The different fluxes
in the displayed light curves cannot be explained by heating alone,
however, and the presence of an accretion-induced component is
indicated for at least the 1992 and 2010 intermediate states.

In a first step toward an ephemeris, we corrected the observed
minimum times in 1992, 1997, and 2010 for this additional light
source. We performed Fourier fits to the light curves, using two
sinusoids with periods $P/2$ and $P$. The former approximates the
ellipsoidal light curve (assuming two equal minima) and the latter
represents the additional light source. The fit has five free
parameters, the period $P$, the two amplitudes, and two phase
shifts. Higher harmonics are not needed for the present purpose. The
resulting fits are shown as solid lines in Fig.~\ref{lc2} (bottom four
panels) and an example of the $P/2$ sinusoid is added as a dotted
curve in the third panel from the bottom. We used these fits for the
sole purpose of correcting the observed minimum times to those that
the $P/2$ component alone would have had. In the low state, this
correction practically vanishes. Table~\ref{tab:monet} lists the
minimum times with this small correction included.
In a second step, we performed a period search around P/2, using the
times of Table~\ref{tab:monet}. The best-fit period is
$P/2\!=\!0.072431961(18)$~d with \,$\chi^2\!=\!22.2$ for 23 degrees of
freedom (d.o.f.). The next-best alias corresponds to a difference of
one cycle on the $P/2$ scale over one year. It has an unacceptable
\,$\chi^2\!=\!58.0$ and can be excluded. Hence, our fit fixes the
orbital period $P$, but still leaves us with the choice of which of
the two minima corresponds to inferior conjunction of the
secondary. It is encouraging though that the deeper minima in 1992 and
2010 are separated by an even number of $P/2$ cycles and thus occur at
the same orbital phase. We adopt the deeper minimum in the
intermediate states as corresponding to inferior conjunction, a choice
that we justify in Sect.~\ref{sec:vrad}, and obtain the ephemeris
\begin{equation}
T_0 = {\rm BJD(TDB)}\ 24\,55215.96256(48) + 0.144\,863\,923\,(36)\,E,
\label{eq:ephem}
\end{equation}
where $T_0$ refers to the inferior conjunction of the secondary
star. All orbital phases quoted in this paper refer to Eq.~1.

\subsection{Orbital variations}

Figures~\ref{lc1} and \ref{lc2} display the X-ray fluxes, optical
fluxes, and radial velocities folded over the orbital period. We
determined the times of maximum flux or of maximum positive radial
velocity by fitting sinusoids to the data, which are shown in the as
dotted or solid curves in the appropriate panels. All time tags from
the optical telescopes and from the ROSAT satellite are given in UTC,
while XMM Newton has the leap seconds added and provides times in
Terrestrial Time (TT). We converted all times to Barycentric Dynamical
Time (TDB), which agrees with TT at the ms
level\footnote{http://www.cv.nrao.edu/$\sim$rfisher/Ephemerides/times.html},
and corrected them for the light travel time to the solar system
barycenter, i.e., we use Barycentric Julian Days in TDB or
BJD(TDB). The conversion from UTC was made with the interactive tool
\texttt{astroutils}\footnote{http://astroutils.astronomy.ohio-state.edu/time/}.
Note that we quote Julian days and not modified Julian days.

\subsubsection{X-ray flux}

The ROSAT X-ray light curves display an orbital maximum near phase
$\phi\!\simeq\!0.78$. A pronounced dip in the light curve of the
pointed ROSAT observation (Fig.~\ref{lc1}, second panel from top)
occurs just before maximum, which lasts from $\phi\!=\!0.73$ to 0.78
and is likely produced by the accretion stream crossing the line of
sight. Ramsay \& Cropper (2003) observed \rxj\ with the CCD cameras on
board of XMM-Newton for part of the orbital period at unknown
phase. With the ephemeris of Eq.~\ref{eq:ephem}, we now find that
their observation intervals covered orbital phases
$\phi\!=\!0.13-0.71$ (EPIC MOS) and $\phi\!=\!0.31-0.69$ (EPIC pn),
just missing the dip. The ROSAT light curves show X-ray emission at
all orbital phases, suggesting either that the X-ray emitting spot
never disappears completely behind the horizon, or that we are seeing
emission from more than one region on the white dwarf. The X-ray flux
varies substantially on time scales down to minutes. Orbit-to-orbit
variability is responsible for the low fluxes in the RASS light curve
(Fig.~\ref{lc1}, top panel, open circles).

\begin {table}[b]
\caption[]{Times and phases of events around the orbit}
\label{tab:events}
\begin {flushleft}
\begin {tabular}{lcc}
\noalign{\smallskip}
\hline
\noalign{\smallskip}
\multicolumn{1}{l}{Maximum of} & 
\multicolumn{1}{c}{BJD(TDB)}  &
\multicolumn{1}{c}{$\phi$} \\ 
& 24000000+ & \\
\noalign{\smallskip}
\hline
\noalign{\smallskip}
X-ray flux RASS 1990             &  48220.1588\,(40)  & 0.76(3) \\
X-ray flux ROSAT 1992            &  48943.6190\,(40)  & 0.83(3) \\
\hb--\heps, HeII radial velocity 1992 &  48631.7244\,(40)  & 0.81(3) \\
\hb--\hg, HeII radial velocity 2000   &  51577.6920\,(50)  & 0.91(4) \\
Visual flux 2000   &  51577.6197\,(50)  & 0.41(4) \\
NaI radial velocity 1997         &  50513.7086\,(70)  & 0.20(5) \\
\ha\ radial velocity 1997        &  50513.7260\,(70)  & 0.32(5) \\
\ha\ line flux 1997              &  50513.7477\,(70)  & 0.47(5) \\
Cyclotron flux 3670\AA\ 1992     &  48630.6984\,(40)  & 0.73(3) \\
Cyclotron flux 3750\AA\ 1997     &  50513.6496\,(30)  & 0.79(2) \\
\noalign{\smallskip}
\hline
\end {tabular}
\end {flushleft}
\end {table}

\subsubsection{Radial velocities in the intermediate and high states}
\label{sec:vrad}

Peak radial velocity of the Balmer and HeII$\lambda4686$ line emission
in the 1992 and 2000 intermediate and high states occur at orbital
phases near X-ray maximum, suggesting that the line emission
originates in the magnetically confined section of the accretion
stream that leads from the stagnation point in the magnetosphere to
the hot spot on the white dwarf
\citep[e.g.][]{schwopeetal00-QQVul,staudeetal01-V1309Ori}. The large mean
radial-velocity amplitudes of the the H$\beta$,
H$\gamma$, H$\delta$, H$\epsilon$, and He II $\lambda$ 4686 emission
lines, $K\!=\!484\pm\!30$\,\kms\ in 1992 and
and $K\!=\!470\pm\!30$\,\kms\ in 2000, support this notion. At this
point, it is appropriate to comment on the phase convention adopted in
Section~\ref{period}. Had we assigned $\phi\!=\!0$ to the less deep
$I$-band minimum, we would face the implausible situation that the
radial velocity maxima would occur half an orbit offset from the X-ray
maxima. Furthermore, the coincidence of maximum X-ray and 1992 and 1997
cyclotron fluxes would be destroyed.

\subsubsection{Optical-continuum fluxes}
\label{sec:emili}

The quasi-$U$ band fluxes in the center two panels of Fig.~\ref{lc1}
represent the cyclotron flux in the third harmonic, with a small
contribution from the white dwarf in 1997 and an additional larger
contribution from the accretion stream in 1992. Peak flux agrees in
phase with the X-ray maximum and the radial velocity maximum in 1992,
and approximately also in 2000. We note that cyclotron beaming, which
shapes the optical light curves in many polars, is not prominent for
the low harmonics observed in \rxj. The quasi-$V$ band flux
$f_\mathrm{vis}$ observed in 2000 displays an entirely different
behavior, being in antiphase to the emission line radial velocities
$v_\mathrm{rad}$ (bottom panels of Fig.~\ref{lc1}).  The high-state
continuum flux must, therefore, have an origin different from the low-
and intermediate-state cyclotron fluxes. A straightforward
interpretation assigns the emission to the accretion stream as the
dominant light source in the high state. The excess flux of the 2MASS
data over the infrared continuum of the secondary star would then
represent the Brackett and Pfund continua emitted in an intermediate or
high state. The observed phase shift between
$f_\mathrm{vis}$ and $v_\mathrm{rad}$ in 2000 can be understood if the
free-free and recombination continuum of the stream is optically thick
and its brightest inner section is hidden from view at the phase of
maximum recessional velocity. This model implies that cyclotron
emission is not a major contributor to the optical flux in the high
state. Cyclotron line emission, while the dominant source of the
low-state $U$-band flux, is restricted to the optically thin fringes
of the accretion spot in the high state and possibly shows little
variation between the low and high states. This behavior is expected
if high-density accreted matter carries its energy into
subsphotospheric layers instead of expanding it in free-standing
shocks above the photosphere of the white dwarf
\citep[e.g.][]{kuijperspringle,fischerbeuermann}.

\subsubsection{Radial velocities and \ha\ line flux in the low state}
\label{sec:emili}

The top three panels in Fig.~\ref{lc2} display the \ha\ line flux in
arbitrary units, the NaI$\lambda8190$ radial velocity, and the \ha\
radial velocity (open circles), as measured from the 1997
low-resolution spectra. The line flux peaks at $\phi\!=\!0.5$, when
the irradiated face of the secondary is in view, and maximum
recessional radial velocity occurs at $\phi\simeq0.25$, suggesting
that these quantities trace the motion of the secondary, providing
additional support for our phase convention. Heating does occur, as
indicated by the fact that the NaI line is best defined when the back
side is in view. However, the \ha\ flux does not drop to zero at
$\phi\!=\!0$, as expected for an origin on the heated side of the
secondary and illustrated by the dashed curve calculated for an
inclination $i\!=\!73^\circ$ (Sect.~\ref{sec:cyc}). Higher resolved
spectra are needed to resolve the origin of the low-state \ha\
emission.
The measured radial-velocity amplitudes are
$K'_\mathrm{2,H\alpha}\!=\!218\pm 30$\,\kms\ and
$K'_\mathrm{2,NaI}\!=\!278\pm 30$\,\kms. The true radial velocity
amplitude $K_2$ of the secondary is probably larger than the former value
and smaller than the latter, and we compromise on
$K_2\!\simeq\!250\!\pm\!20$\,\kms.

\subsection{The secondary star}
\label{sec:sec}

\begin{figure}[t]
  \includegraphics[height=8.9cm,bb=85 67 545 699,angle=270,clip=]{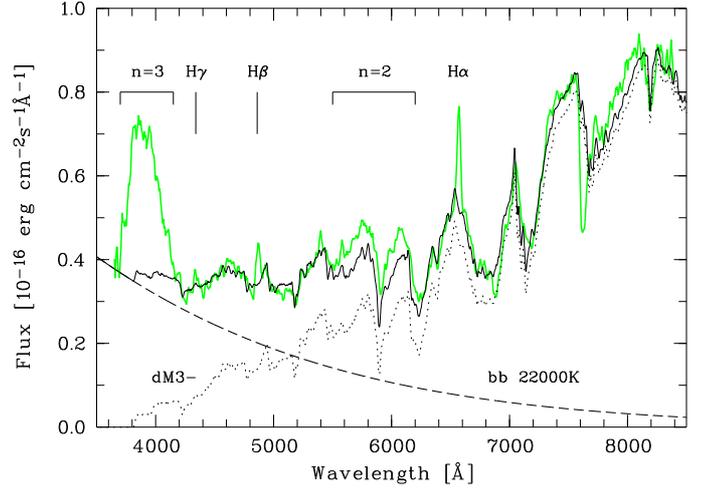}
  \caption{\label{fig:sec} Observed spectrum in the 1997 low state at
    cyclotron minimum (green curve), fitted by a dM3- star and a
    blackbody (see text). }
\end{figure}

TiO features of the secondary star dominate the 1997 low-state spectra
longward of 6200\,\AA. The secondary star is best detected at orbital
minimum of the cyclotron flux near $\phi\!=\!0.25$, which corresponds
to the maximum of the ellipsoidal modulation. The mean of the three
spectra at $\phi\!=\!0.18$, 0.26, and 0.34 (Fig.~\ref{fig:sec}, green
curve) shows the M star, the blue continuum of the white dwarf, and
remnant cyclotron emission in the restricted intervals of
3700--4200\,\AA\ (third harmonic) and 5500-6200\,\AA\ (second
harmonic).
We fitted the observed spectrum in the intervals free of cyclotron
emission, using representative spectra of M2 to M4 dwarfs from the
Sloan Digital Sky Survey \citep{rebassaetal12} and a blackbody for the
white dwarf. We find a spectral type dM3-- with an uncertainty of half
a spectral class (dotted black curve) and a blackbody temperature of
22000\,K (dashed black curve) for a white dwarf of 0.8\,\msun\ with a
radius of $\ten{7.4}{8}$\, cm at a distance of 790\,pc
(Sections~\ref{sec:masses} and \ref{sec:dist}). Extinction with
$A_\mathrm{I}\!\sim\!0.04$, as suggested by the interstellar absorbing
hydrogen column density derived from the X-ray spectral fits
(Section~\ref{sec:x}), implies an $I$-band flux higher by 4\%. The
7500\,\AA\ spectral flux of the secondary reduced to the level of the
primary minimum at $\phi\!=\!0$ (Fig.~\ref{lc2}, third panel from
bottom) then is $f_{7500}=\ten{6.6}{-17}$\,\ergsa\ with an error of
10\% based on the uncertainties in the flux calibration. Differential
extinction would lead to a minimally earlier spectral type than
adopted above. This value of $f_{7500}$ is used in
Sect.~\ref{sec:dist} to derive the distance of \rxj.

The amplitude of the ellipsoidal modulation is 25\% of the peak flux,
a number typical of a dM star viewed at high inclination $i$. For a
dM3-- star and $i\!=\!73^\circ$ (Sect.~\ref{sec:cyc}), the observed
modulation amplitude is reproduced using the appropriate
gravity-darkening coefficient $\beta_1\!=\!0.65$ and limb-darkening
coefficient $u_\mathrm{l}\!=\!0.216$ from \citet{clareta,claretb}.
The ellipsoidal modulation is also seen in the wavelength interval
$\lambda\!=\!4400\!-\!5200$\,\AA, where the secondary still
contributes about half of the observed flux.

\subsection{Masses of the stellar components of \rxj}
\label{sec:masses}

The mean Roche-lobe radius of the secondary star is given by
\begin{equation} 
R_2/R_{\odot} = 0.234\,(M_2/M_{\odot})^{1/3}\,P^{2/3}f,
\label{eq:radius}
\end{equation} 
\noindent where $M_2$ is its mass, $P$ is the orbital period in hours,
and $f\!\simeq\!0.989$ (Kopal 1959). In CVs above the period gap, mass
loss has usually driven the secondary somewhat out of thermal
equilibrium \citep{kniggeetal11}. As a consequence, its radius is
increased beyond that of an undisturbed field star by a higher
percentage than expected from tidal and centrifugal forces alone
\citep{kippenhahnthomas70,renvoizeetal02}.
Depending on the long-term average of the mass transfer rate $\dot{M}$,
the secondary could have a mass as large as 0.40\,\msun\ with a
spectral type dM2.9 or as low as 0.24\,\msun with a spectral type
dM3.9 \citep[][their Table 2]{kniggeetal11}. The observed spectral
type of dM3- favors a larger mass and settle on $M_2\!=\!0.35\pm
0.05$, which allows for moderate bloating. From Eq.~\ref{eq:radius}, we
obtain $R_2\!=\!(2.60\pm\!0.13)\,10^{10}$\,cm.
An estimate of the primary mass $M_1$ can be obtained from the
radial-velocity amplitude $K_2$ derived in Section~\ref{sec:emili}.
With $i\!=\!73^\circ$, we obtain $M_1\!=0.80\!\pm\!0.15$\,\msun.
 
\subsection{The distance of \rxj}
\label{sec:dist}

We obtained the distance of \rxj\ from the observed spectral flux of the
secondary star combined with its radius and surface brightness.
Roche geometry yields
an effective radius of the projected cross section of the star at
$\phi\!=\!0$ and $i\!=\!73^\circ$ of
$R_\mathrm{proj}\!=\!0.957\,R_2\!=\!\ten{2.49}{10}$\,cm.
With the 7500\,\AA\ spectral flux of the secondary star from above,
$f_{7500}=\ten{(6.6\!\pm\!0.7)}{-17}$\,\ergsa, and the surface
brightness of a dM3-- star at 7500\,\AA\ of
$F_{7500}=\ten{(6.4\!\pm\!0.9)}{5}$\,\ergsa\ \citep{beuermann06}, the
distance is obtained as
$d\!=\!R_\mathrm{proj}\,(F_{7500}/f_{7500})^{1/2}\!=\!\ten{(2.45\!\pm\!0.32)}{21}\,\mathrm{cm}=790\!\pm\!105$\,pc.

\begin{figure}[t]
  \includegraphics[width=89mm,bb=93 506 468 735,clip=]{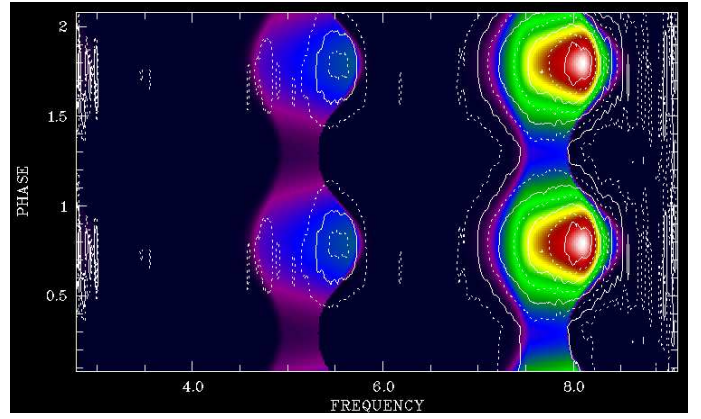}
  \caption{2-D representation of the phase-dependent observed and
    calculated cyclotron lines of \rxj\ in 1997. The best-fit
    calculated spectral flux (color) is compared with the observed
    fluxes (contour lines). The abscissa is frequency in units of
    $10^{14}$\,Hz, the ordinate is phase and data and model are
    displayed twice. The image shows the second and third harmonic
    with an indication of the first harmonic at the lowest
    frequencies. }
\label{fig:cyc2} 
\end{figure}

\subsection{Cyclotron spectroscopy}
\label{sec:cyc}

Cyclotron line emission dominates the 1997 low-state spectra and
displays a marked orbital variation with a minimum at
$\phi\!\simeq\!0.3$ and a maximum at $\phi\!\simeq\!0.8$. The latter
coincides with the X-ray maximum (Fig.~\ref{lc1}). Peak flux occurs
when the line of sight is closest to the axis of the accretion
funnel. This differs from the situation encountered in low-field polars,
where cyclotron beaming results in a flux minimum in the direction of
the funnel. In \rxj, however, we observed in low harmonics with little
or no preference of emission perpendicular to the field (note that the
first harmonic is emitted primarily along the field).

\begin{figure}[t]
\includegraphics[height=8.9cm,bb=248 67 491 705,angle=270,clip=]{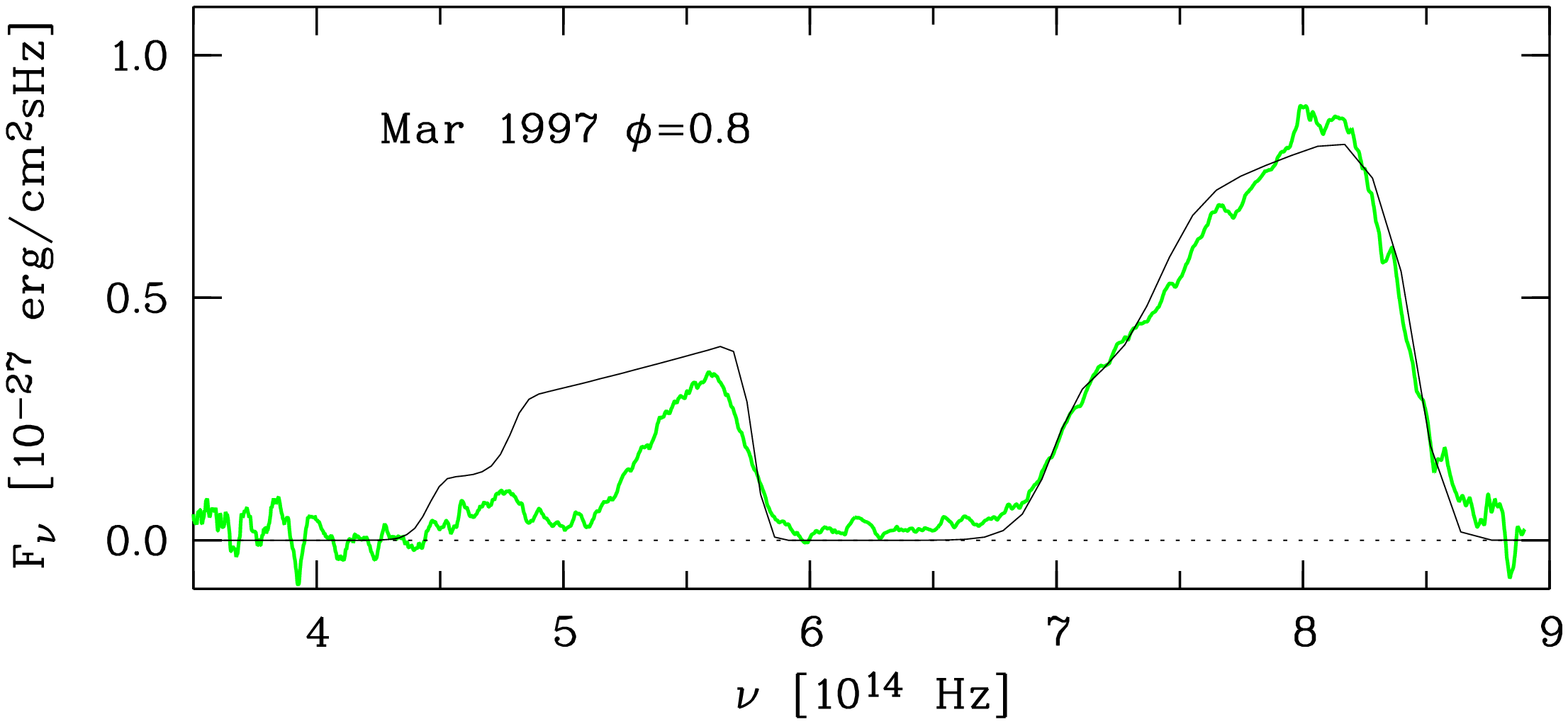}
  \includegraphics[height=89mm,bb=248 67 543 705,angle=270,clip=]{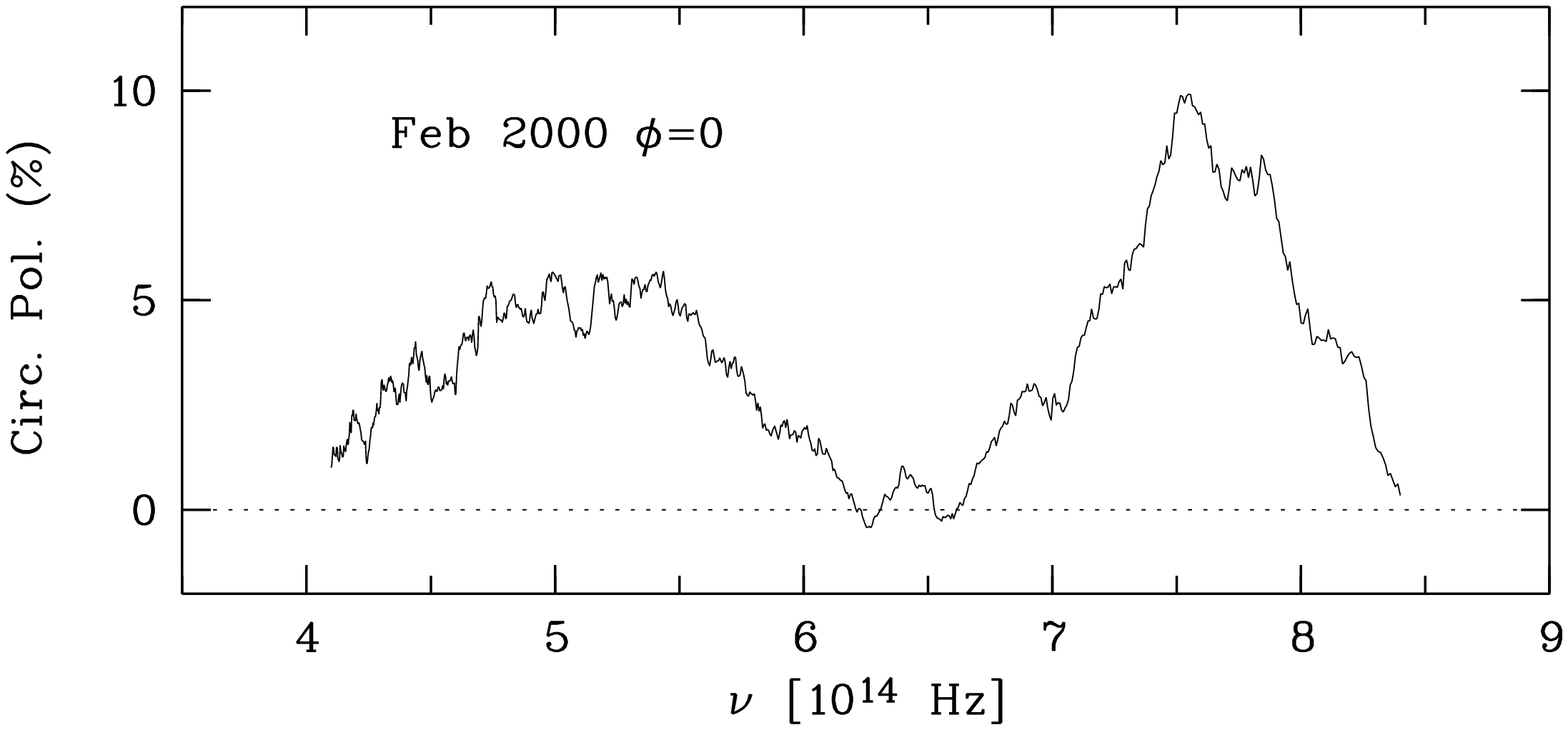}
\caption{Representative cyclotron spectra showing the second and third
  harmonic in a field of 94\,MG. \emph{Top: }Cyclotron spectrum at
  $\phi\!=\!0.8$ in the 1997 low state (green) and
  theoretical cyclotron spectrum (black) obtained from a
  simultaneous fit to the entire set of spectra as shown in
  Fig.~\ref{fig:cyc2}.  \emph{Bottom: }Circular polarization in the high
  state on 6/7 February 2000 measured around the flux minimum at
  orbital phase $\phi\!=\!0$.}
\label{fig:cyc1}
\end{figure}

The phase-resolved cyclotron spectra were obtained by subtracting the
contributions of the secondary and primary star from the 13 observed
spectra of the 1997 data set.  At peak flux, the second and third
harmonics extend over the wavelength intervals
5200--6700\,\AA\ and 3500--4400\,\AA,
respectively. Between these intervals, the cyclotron flux is
effectively zero. For a more detailed analysis, we rebinned the spectra to the
spectral resolution of 50\,\AA\ and assigned 'errors' to the individual
spectral bins that reflect the scatter of the original data points
within each 50\,\AA\ bins. These spectra are displayed on a
frequency scale in the 2-D image of Fig.~\ref{fig:cyc2} (contour
lines). As an example, a single spectrum at maximum cyclotron flux is
shown in Fig.~\ref{fig:cyc1} (top panel, green curve). 

\begin {table}[b]
\caption[]{Best-fit parameters for the fit shown in Figs.~\ref{fig:cyc2}
  and ~\ref{fig:cyc1}.}
\label{tab:cyc}
\begin {flushleft}
\begin {tabular}{lrc}
\noalign{\smallskip}
\hline
\noalign{\smallskip}
\multicolumn{1}{l}{Parameter} & 
\multicolumn{1}{@{\hspace{10mm}}c}{Value} & 
\multicolumn{1}{c}{Error} \\
\noalign{\smallskip}
\hline
\noalign{\smallskip}
Magnetic field strength (MG) & 94.0 & 0.3 \\
Plasma temperature (keV)) & 2.1 & 0.3 \\
Size Parameter $\Lambda$ ($10^4$) & 2.8 & \hspace{3.4mm}+2.3,--1.3 \\
Solid angle $\Omega$ ($10^{-26}$\,sr) & 3.0 & 0.3 \\
Inclination $i$ ($^\circ$) & 73\hspace{2.3mm} & 1\hspace{2.3mm} \\
Colatitude of field/spot $\beta$ ($^\circ$) & 12.3 & 1.6 \\
Azimuth of field/spot  $\psi$ ($^\circ$) & 70\hspace{2.3mm} & 4\hspace{2.3mm} \\
\noalign{\smallskip}
\hline
\end {tabular}
\end {flushleft}
\end {table}

We fitted the entire set of cyclotron spectra simultaneously, using
model spectra calculated for an isothermal homogeneous plasma slab \citep
[][so-called constant-$\Lambda$ model]{barrett}. The free parameters
of the multi-parameter fit were the magnetic field strength \(B\), the
plasma temperature \(T\), the dimensionless slab thickness
\(\Lambda\), the solid angle $\Omega$ subtended by the accretion spot
as seen from the Earth, the inclination \(i\) of the system, and the
direction of the accreting field line given by the polar angle
\(\beta\) and the the azimuth \(\psi\). The location of the spot is
assumed to agree with the foot-point of the field line. We found that a
consistent set of parameters exists that simultaneously fits the
thirteen spectra with 50 data points each. A color representation of
the calculated spectral flux is shown in Fig.~\ref{fig:cyc2}. The spectra
represent the second and third harmonics in a field of 94\,MG.
The complete set of fitted parameters is provided in
Table~\ref{tab:cyc} along with their $1\!-\!\sigma$ errors. The fit
has a \,$\chi^2\!=\!517$ for 643 d.o.f.  It is remarkable that a
close-to-perfect fit to the observed data set is possible using the
fairly simple theoretical model. The fit fixes the orbital inclination
of \rxj\ at $i\!=\!73^\circ\pm 1^\circ$ ($1\!-\!\sigma$ error),
implying that the system just escapes eclipse of the white dwarf by
the secondary star. With 2.1\,keV, the plasma temperature is lower
than the mean temperature of 6\,keV derived from the fit to the XMM
spectrum, but both temperatures are substantially lower than expected
for a free-standing shock on a white dwarf of 0.8\,\msun. 

\begin{figure*}[t]
  \includegraphics[height=92.5mm,bb=83 83 543 602,angle=270,clip=]{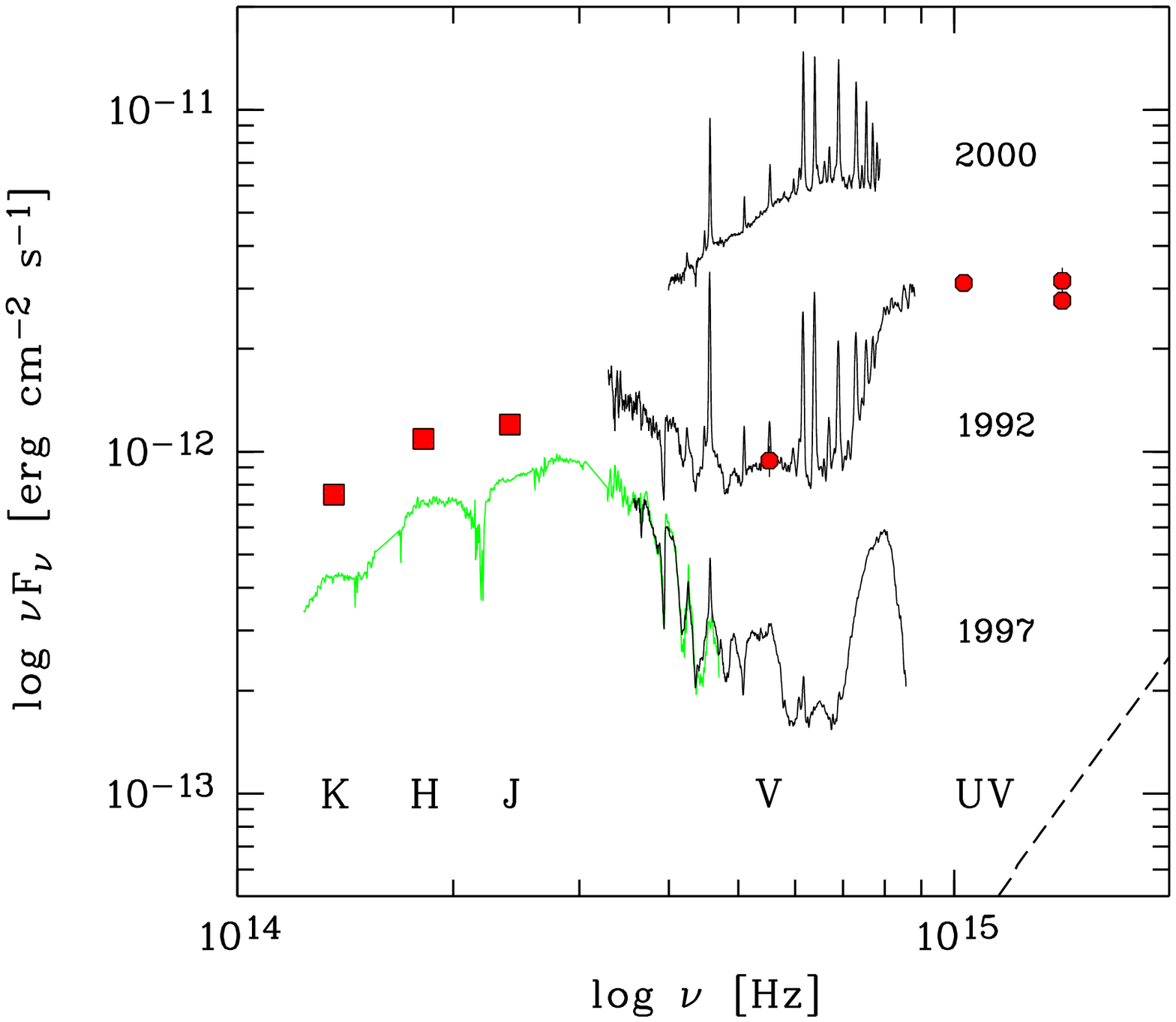}
\hfill
  \includegraphics[height=87.0mm,bb=80 34 543 524,angle=270,clip=]{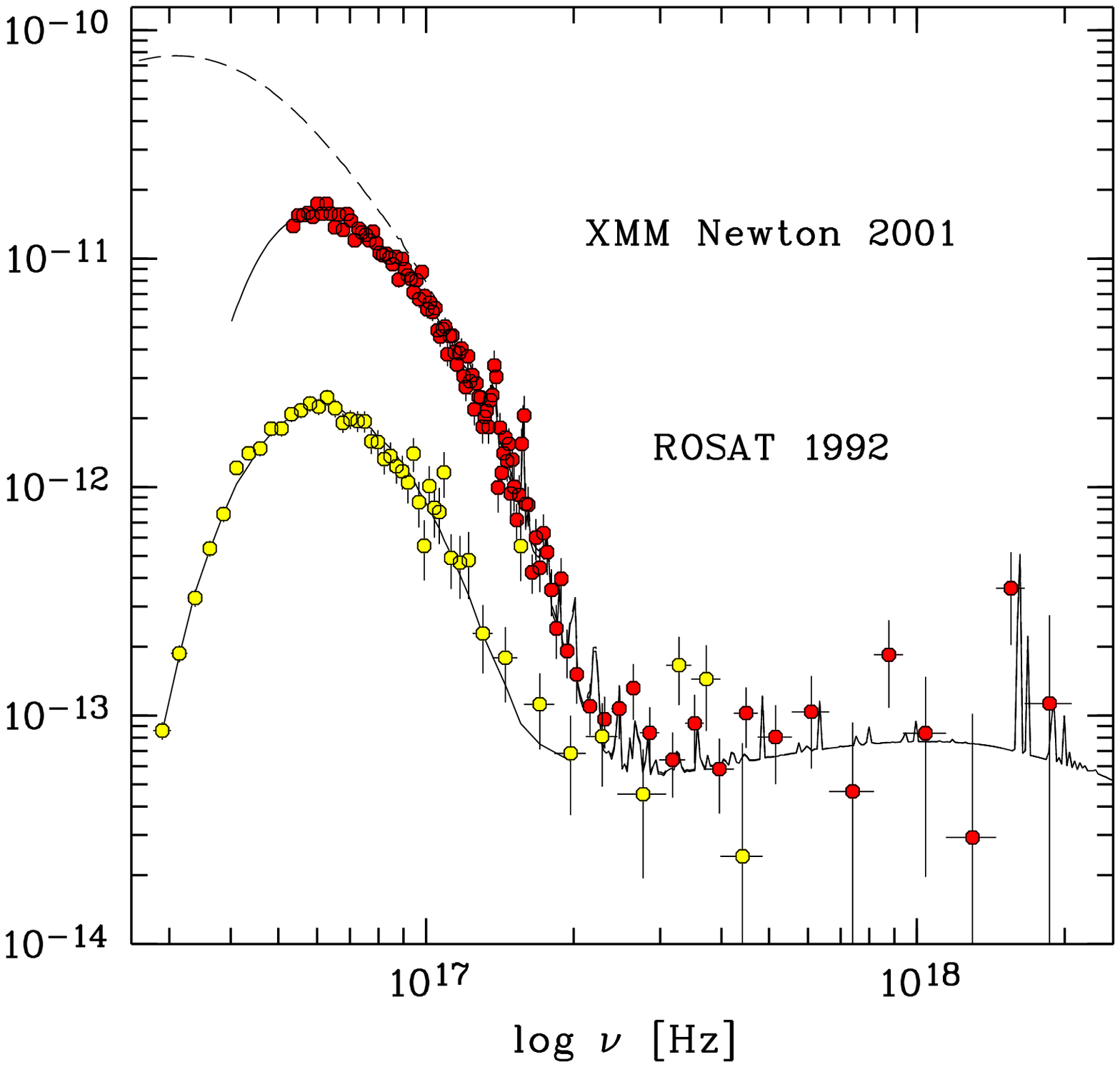}
  \caption{ Spectral energy distribution of RXJ1007-20. \emph{Left: } IR-optical-UV part. \emph{Right: } X-ray part. See text for further explanation. }
\label{fig:sed}
\end{figure*}

The dominance of emission from the magnetic pole with $B\!=\!94$\,MG
persists from the low state into the high state of \rxj.  Even against
the intense Paschen continuum of the 2000 high state, the second and
third harmonics stand out in the circular polarization at practically
the same wavelengths as in the low state (Fig.~\ref{fig:cyc1}, bottom
panel). The line shape, furthermore, suggests a low temperature
similar to the 2\,keV of the low state, consistent with emission from
the low-density fringes of the accretion spot, in which cyclotron
emission is the prime cooling agent (Woelk \& Beuermann 1996, Fischer
\& Beuermann 2001).  Radiation-hydrodynamic calculations suggest that
columns with different temperatures and plasma densities may coexist
in the spot and that cyclotron cooling dominates in low-density
columns, in which the plasma cools before the ion and electron
temperatures can equilibrate (Woelk \& Beuermann 1996, Fischer \&
Beuermann 2001).

Despite the generally excellent agreement, some differences between
the observed and calculated line profiles are noteworthy. An observed
feature that is not present in the blackbody-limited line flux is the
pronounced central depression in the second harmonic around
6000\,\AA. It is likely that this depression represents the
Fraunhofer-type absorption produced by the temperature stratification
in columns that are optically thick in the emitted cyclotron harmonics
\citep{woelketal96}. A similar profile was observed in the cyclotron
lines of the high-field polar UZ~For \citep{rousseauetal96} and was
successfully modeled with the theory of \citet{woelketal96}. The
slight notch in the long-wavelength fall-off of the model profiles
results from differences in the line profiles for the ordinary and
extraordinary rays in this fairly simple theory.

\subsection{The X-ray spectrum}
\label{sec:x}

\begin{table}[b]
\caption{Bolometric fluxes and luminosities of RXJ1007 for the
  intermediate states of 1992/2001. See text for the geometry factors $\eta\pi$.}
\label{tab:sed}
\begin{tabular}{l@{\hspace{4.0mm}}c@{\hspace{4.0mm}}c@{\hspace{4.0mm}}c}
\hline \hline \noalign{\smallskip}
Component & F           & $\eta \pi$ & $L$    \\           
          & (\ergs)     &            & (\erg)    \\           
\noalign{\smallskip} \hline
\noalign{\smallskip}
Soft X-ray blackbody,\hspace{2.4mm}32\,eV    & $8.5\times 10^{-11}$  & $2\,\pi$   & $3.2\times 10^{33}$  \\
\hspace{30.3mm}59\,eV   & $3.5\times 10^{-11}$  & $2\,\pi$   & $1.3\times 10^{33}$  \\
Thermal X-rays\hspace{8.0mm}0.2\,keV & $9.0\times 10^{-13}$ & $4\,\pi$ & $6.7\times 10^{31}$  \\
\hspace{28.3mm}6.0\,keV    & $2.8\times 10^{-13}$ & $4\,\pi$ & $2.1\times 10^{31}$  \\  
\hspace{26.2mm}$>10$\,keV ?& $1.5\times 10^{-13}$ & $4\,\pi$ & $1.1\times 10^{31}$  \\
Cyclotron + accretion stream      & $4.0\times 10^{-12}$ & $4\,\pi$   & $3.0\times 10^{32}$  \\[0.6ex]      
Total      &   & & $4.9\times 10^{33}$  \\[1.0ex]      
\noalign{\smallskip} \hline  \noalign{\smallskip}
\end{tabular}
\end{table}

We have re-analyzed the mean X-ray spectrum taken with XMM-Newton on 7
December 2001 \citep[][their Fig.~8]{ramsaycropper03}, using an
improved spectral representation of the soft and hard X-ray
components. As demonstrated for AM Her \citep{beuermannetal12}, a
model including two or more blackbody components may yield a more
realistic estimate of the soft X-ray luminosity than the
single-blackbody assumption employed by \citet{ramsaycropper03}. Our
fit involves two blackbodies with best-fit temperatures
k$T_1\!=\!59\pm3$\,eV and k$T_1\!=\!32\pm5$\,eV for a best-fit
absorbing column density $N_\mathrm{H}\!=\!\ten{1.9}{20}$\,\atoms. The
hard X-ray component was modeled with two MEKAL spectra for solar
abundances and temperatures k$T\!=\!6.0$\,keV (fixed) and
k$T\!=\!0.19\pm0.01)$\,keV. The fit has a \,$\chi^2\!=\!115$ for
107\,d.o.f. (\,$\chi^{2}_\nu\!=\!1.07$). The integrated fluxes for the
best-fit parameters are $\ten{3.5}{-11}$\,\ergs\ and
$\ten{8.5}{-13}$\,\ergs\ for the two blackbodies and
$\ten{3.5}{-13}$\,\ergs\ and $\ten{8.5}{-13}$\,\ergs\ for the two
thermal components, respectively. Guided by the case of AM~Her
\citep{christian00}, we considered that the hard X-ray component of
\rxj\ possibly contains highly absorbed components that are not
accounted for by the XMM spectral fit at $E\!<\!8$keV. We estimated
that this could contribute another $\ten{(1\!-\!2)}{-13}$\,\ergs. Since
the presence of a substantial hard X-ray component is uncertain, it is
entered in Table~\ref{tab:sed} with a question mark. We caution,
furthermore, that the correlated errors between the blackbody
temperature and the interstellar absorbing column imply large
uncertainties, preventing us from quoting errors for the
fluxes. Despite all the uncertainties, however, the
emission of \rxj\ is clearly dominated by soft X-rays. In the 1992 pointed
ROSAT observation, the source was fainter by almost an order of
magnitude. The fit parameters are less well-defined than those of the
XMM fit and are not quoted here.

\subsection{Overall spectral energy distribution}
\label{sec:sed}

Fig.~\ref{fig:sed} shows the spectral fluxes from the infrared to the
hard X-ray regime collected into a single overall spectral energy
distribution (SED). For the chosen form $\nu f_{\nu}$ vs. $\nu$, the
integrated energy flux of a component with spectral flux $f_{\nu}$ is
$F\!=\!\int\nu f_{\nu}~\mathrm{dlog}\,\nu$, with log the natural
logarithm.  Given the flux of a spectral component, the luminosity
\mbox{$L\!=\!\eta \pi d^2F$} was calculated with $d\!=\!790$\,pc and
the geometry factor $\eta$ quoted in Table~\ref{tab:sed}. As for the
prototype polar AM~Her \citep{beuermannetal12}, we adopted emission of
the soft X-ray blackbody and of cyclotron radiation into $2\,\pi$ and
of the accretion stream and the hard X-rays into $4\,\pi$.

The left panel shows our spectrophotometry of 1992, 1997, and 2000
(black curves), supplemented by the 2MASS J, H, and K-band fluxes of
1999 (red filled squares) and the visual and ultraviolet fluxes
measured with the optical monitor on board of XMM-Newton simultaneous
to the 2001 X-ray observation (red filled circles). For the low state,
which is dominated by the secondary star, we added the flux
distribution of the dM3 star LHS58 \citep[][green
  curve]{leggettetal96} adjusted to the 1997 low-state
spectrophotometry. The right panel shows the incident spectra for the
2001 XMM pn observation (red filled circles) and the 1992 ROSAT PSPC
observation (open circles). For the XMM spectrum, we also
included the 'source spectrum' corrected for interstellar absorption,
using the X-ray spectral parameters quoted above (dashed curve in the
left and the right panel).

There is a close agreement between our 1992 spectrophotometry and the
2001 XMM optical-monitor fluxes in the visual and ultraviolet bands,
suggesting that they delineate the common SED of an intermediate state
of accretion assumed by \rxj in 1992 and 2001. This SED is well
documented from the infrared to the hard X-ray regime. About 92\% of
the bolometric flux in this state is emitted as soft X-rays, 6\% as
cyclotron radiation and stream emission, largely in the ultraviolet,
and only 2\% as hard X-rays. The dominance of soft X-rays implies that
most of the accretion energy is released in shocks, which are buried
in the photosphere of the white in a scenario first described by
\citet{kuijperspringle}. The luminosity ratio
$L_\mathrm{softX}/(L_\mathrm{cyc}+L_\mathrm{hardX}+L_\mathrm{stream})\!\sim\!11$.
The low-level cyclotron emission in 1997 suggests that the associated
X-ray flux was low as well. Similarly, we can only speculate that the
optical high state in February 2000 was accompanied by a
correspondingly increased soft X-ray flux. The rough equality of the
ROSAT and XMM X-ray fluxes in the 1-2\,keV range hints at a lower
variability of the hard X-ray fluxes, consistent with the notion of
\citet{kuijperspringle} that an increased $\dot M$ arises primarily
from dense blobs, which penetrate to subphotospheric layers.

For a white dwarf with $M_1\!=\!0.8$\,\msun\ and
$R_1\!=\!\ten{7.4}{8}$\,cm, the accretion luminosity in the 1992/2001
intermediate state of $L_\mathrm{acc}\!\simeq\!\ten{4.9}{33}$\,\ergs\
requires an accretion rate $\dot
M\!=\!R_1\,L_\mathrm{acc}/(G\,M_1)\!\simeq\!\ten{3.4}{16}$\,\gs\ or
$\ten{5.4}{-10}\,M_\odot\,{\rm yr}^{-1}$. The lack of X-ray coverage
prevents an estimate of the accretion rate in a high state, but if it
follows the optical continuum, it could exceed
$10^{-9}\,M_\odot\,{\rm yr}^{-1}$ and, thereby, reach into the realm typical of
long-period CVs \citep{kniggeetal11}.

\section{Conclusions}

We have presented a comprehensive study of the high-field polar \rxj,
which includes observations performed over more than 20 years. It is a
highly variable source that was encountered at optical flux levels
differing by 4 mag over the years. In the intermediate state of 1992
and 2001, \rxj\ has a soft X-ray luminosity
$L_\mathrm{softX}\!\ga\!0.90\,L_\mathrm{acc}\!\simeq\!1.2\,L_\odot$. In
a high state of accretion, the soft X-ray luminosity is possibly
higher still, both relatively and absolutely. With a polar field
strength of 94\,MG, \rxj\ belongs to the few high-field
polars. The main accreting pole is permanently in view and is the
source of the cyclotron lines observed at all accretion levels. The
observed cyclotron line profiles and the low temperature in the
emission region are perfectly consistent with cyclotron theory.

\begin{acknowledgements}
  A draft of this paper was left by the late Hans-Christoph Thomas,
  who had finished analysis of the photometric MONET/N data of \rxj\
  collected in early January 2012, before he suddenly died of heart
  failure on 18 January 2012, aged 75. Hans-Christoph Thomas was
  highly regarded as a collaborator and as a friend by many colleagues
  in the communities dealing with stellar structure theory and with
  close binaries. Being one of the earliest collaborators of Rudolf
  Kippenhahn, Hans-Christoph Thomas was the first to follow the
  evolution of a solar-like star numerically through the helium flash.
  Many young scientists benefitted from his ever selfless and friendly
  cooperation.

  This work is based in part on data obtained with the MOnitoring
  NEtwork of Telescopes (MONET), funded by the Alfried Krupp von
  Bohlen und Halbach Foundation, Essen, and operated by the
  Georg-August-Universit\"at G\"ottingen, the McDonald Observatory of
  the University of Texas at Austin, and the South African
  Astronomical Observatory. The spectroscopic observations were
  obtained with the ESO 3.6-m telescope under Program No. 64.H-0311(A)
  and with the ESO/MPI 2.2-m telescope in MPI time.  This paper
  makes use of observations with the X-ray telescopes on board of
  the German/British/American ROSAT satellite and XMM-Newton, an ESA
  Science mission with instruments and contributions directly funded
  by ESA Member States and NASA.
\end{acknowledgements}

\bibliographystyle{aa}

\begin{thebibliography}{29}
\expandafter\ifx\csname natexlab\endcsname\relax\def\natexlab#1{#1}\fi


\bibitem[Baraffe et al.(1998)]{baraffe98} Baraffe I., Chabrier G.,
  Allard F., Hauschild P.H., 1998, A\&A 337, 403

\bibitem[Barret \& Chanmugam(1985)]{barrett} Barrett P.E., Chanmugam
  G., 1985, ApJ 298, 743

\bibitem[Beuermann(2006)]{beuermann06} Beuermann, K.\ 2006, \aap, 460, 783

\bibitem[Beuermann et al.(2012)]{beuermannetal12} Beuermann, K.,
  Burwitz, V., \& Reinsch, K.\ 2012, \aap, 543, A41

\bibitem[Beuermann \& Thomas(1990)]{bt90} Beuermann K., Thomas H.-C.,
  1990, A\&A 230, 326

\bibitem[Beuermann \& Thomas(1993)]{bt93} Beuermann K., Thomas H.-C.,
  1993, Adv. Space Res. 13 (12), 115

\bibitem[Christian(2000)]{christian00} Christian, D.~J.\ 2000, \aj,
119, 1930

\bibitem[Claret(2000a)]{clareta} Claret A., 2000, A\&A 359, 289

\bibitem[Claret(2000b)]{claretb} Claret A., 2000, A\&A 363, 1081

\bibitem[Fischer \& Beuermann(2001)]{fischerbeuermann} Fischer, A., \&
  Beuermann, K.\ 2001, \aap, 373, 211

\bibitem[Fouqu{\'e} et al.(2000)]{fouqueetal00} Fouqu{\'e}, P.,
  Chevallier, L., Cohen, M., et al.\ 2000, \aaps, 141, 313

\bibitem[Kippenhahn \& Thomas(1970)]{kippenhahnthomas70} Kippenhahn,
  R., \& Thomas, H.-C.\ 1970, IAU Colloq.~4: Stellar Rotation, 20

\bibitem[Knigge et al.(2011)]{kniggeetal11} Knigge, C., Baraffe, I., 
\& Patterson, J.\ 2011, \apjs, 194, 28 

\bibitem[Kuijpers \& Pringle(1982)]{kuijperspringle} Kuijpers, J.,
  \& Pringle, J.~E.\ 1982, \aap, 114, L4

\bibitem[Leggett et al.(1996)]{leggettetal96} Leggett, S.~K., Allard, 
F., Berriman, G., Dahn, C.~C., \& Hauschildt, P.~H.\ 1996, \apjs, 104, 117 


\bibitem[Ramsay \& Cropper(2003)]{ramsaycropper03} Ramsay, G., \&
  Cropper, M.\ 2003, \mnras, 338, 219

\bibitem[Rebassa-Mansergas et al.(2012)]{rebassaetal12}
  Rebassa-Mansergas, A., Nebot G{\'o}mez-Mor{\'a}n, A., Schreiber,
  M.~R., et al.\ 2012, \mnras, 419, 806

\bibitem[Reinsch et al.(1999)]{reinschetal99} Reinsch, K.,
  Burwitz, V., Beuermann, K., \& Thomas, H.-C.\ 1999, Annapolis
  Workshop on Magnetic Cataclysmic Variables, 157, 187


\bibitem[Renvoiz{\'e} et al.(2002)]{renvoizeetal02} Renvoiz{\'e}, V.,
  Baraffe, I., Kolb, U., \& Ritter, H.\ 2002, \aap, 389, 485

\bibitem[Rousseau et al.(1996)]{rousseauetal96} Rousseau, T., Fischer,
  A., Beuermann, K., \& Woelk, U.\ 1996, \aap, 310, 526


\bibitem[Schwarz et al.(2005)]{schwarzetal05} Schwarz, R.,
  Reinsch, K., Beuermann, K., \& Burwitz, V.\ 2005, \aap, 442, 271


\bibitem[Schwope et al.(2000)]{schwopeetal00-QQVul} Schwope, A.~D., 
Catal{\'a}n, M.~S., Beuermann, K., et al.\ 2000, \mnras, 313, 533 

\bibitem[Schwope et al.(2009)]{schwope09} Schwope, A.~D., Nebot
  Gomez-Moran, A., Schreiber, M.~R., G\"ansicke, B.~T.\ 2009, \aap,
  500, 867

\bibitem[Staude et al.(2001)]{staudeetal01-V1309Ori} Staude, A.,
  Schwope, A.~D., \& Schwarz, R.\ 2001, \aap, 374, 588

\bibitem[Thomas \& Beuermann(1998)]{tb98} Thomas H.-C., Beuermann K.,
  Reinsch K., et al., 1998, A\&A 335, 467



\bibitem [von Zeipel(1924)]{vzeipel} von Zeipel, H., 1924,
  Mon. Not. Roy. Astron. Soc. 84, 665

\bibitem[Woelk 
\& Beuermann(1996)]{woelketal96} Woelk, U., \& Beuermann, K.\ 1996, \aap, 306, 232 

\end{thebibliography}

\end{document}